\documentstyle[12pt,a4,amssymb,times]{article}

\def\vphi{\varphi}
\def\D{\Delta}
\def\e{\epsilon}
\def\L{\Lambda}
\def\r{\rho}
\def\c{\chi}
\def\g{\gamma}
\def\o{\omega}
\def\p{\pi}
\def\d{\delta}
\def\m{\mu}
\def\n{\nu}
\def\l{\lambda}

\def\pa{\partial}
\def\to{\rightarrow}
\newcommand{\be}{\begin{equation}}
\newcommand{\ee}{\end{equation}} 
\newcommand{\bea}{\begin{eqnarray}}
\newcommand{\eea}{\end{eqnarray}}

\begin{document}

\begin{titlepage}

\begin{flushright} 
October 1999
\end{flushright}

\bigskip
\bigskip
\bigskip
\bigskip

\begin{center}

{\bf{\Large
    Free-field realization of $D$-dimensional cylindrical \\ 
               gravitational waves}}

\bigskip 

 A. Mikovi\'c \footnote{E-mail address: amikovic@ualg.pt. On leave of absence 
from Institute of Physics, P.O.Box 57, 11001 Belgrade, Yugoslavia}
and N. Manojlovi\'c \footnote{E-mail address: nmanoj@ualg.pt}

\end{center}

\begin{center}

\footnotesize
	 \'Area Departamental de Matematica, UCEH, Universidade do Algarve,
         Campus de Gambelas, 8000 Faro, Portugal 
\end{center}

\normalsize 

\bigskip 
\bigskip
\begin{center}
			{\bf Abstract}
\end{center}	
We find two-dimensional free-field variables for $D$-dimensional 
general relativity on spacetimes with $D-2$ commuting 
spacelike Killing vector fields and non-compact spatial sections for $D> 4$. 
We show that there is a canonical transformation which maps 
the corresponding two-dimensional dilaton gravity theory into a    
two-dimensional diffeomorphism invariant theory of the free-field variables. 
We also show that the spacetime metric components can be expressed as
asymptotic series in negative powers of the dilaton, with coefficients which 
can be determined in terms of the free fields.    

\end{titlepage}
\newpage

\section{Introduction}

Symmetry reductions of higher-dimensional gravity theories 
are interesting for several
reasons. One is that new classical solutions and new
integrable models can be obtained. Furthermore, these solutions 
can be also interpreted via 
dimensional reduction as solutions for $D=4$ general
relativity with matter.
Second, the quantization of such systems can be helpful for understanding 
conceptual problems of quantum gravity \cite{K,A,nk}, as well as for
understanding issues related with quantum properties of black holes
\cite{strom}.
And third, new mathematical structures arise.

Particularly interesting are symmetry reductions which give $D=2$ integrable
models, because all the relevant issues can be more easily explored. 
Free-field realizations of $D=2$ integrable models are extremely useful
for understanding the structure of the space of solutions and especially 
for the quantization of the theory. One can obtain them via Backlund 
transformations, as in the case of the Liouville model \cite{liu}, or 
via free-field realizations of the symmetry
algebras \cite{frenk}. In the context of $D=2$ integrable models of
gravity, free-field realizations have been found for many models
\cite{mik,CJZ,Cruz,cmn} and this has been explained in \cite{cmn} as the
consequence of the integrability and special properties of the $D=2$ 
diffeomorphism algebra of the constraints.

In this paper we will consider $D$-dimensional general relativity, with $D>4$,
on spacetimes with $D-2$ commuting spacelike Killing
vector fields. In $D=4$, this system corresponds to cylindrically symmetric
general relativity. The corresponding dynamical systems are exactly 
integrable, and the integrability for $D>4$ follows
trivially from the proof of integrability in the $D=4$ case in the
Belinski-Zakharov-Maison (BZM) approach \cite{bz,m}. One 
simply replaces the relevant two by two matrices with $D-2 \times D-2$
matrices. The symmetry reduced theory is a $D=2$ dilaton gravity coupled to a
coset space sigma model \cite{bmg}. In the $D=4$ case when the spatial 
section is non-compact, which corresponds to cylindrical gravitational waves, 
it has been shown in \cite{cmn} that there is a canonical transformation 
which maps the constraints of the symmetry reduced theory
into a free-field form, so that the initial theory is equivalent to a 
two-dimensional
diffeomorphism invariant theory of four free fields. Since $D>4$ case involves
only bigger matrices, it is reasonable to expect that the free-field 
construction of \cite{cmn} could be also generalized, so that the $D>4$ 
reduced theory should be
equivalent to a two-dimensional diffeomorphism invariant theory of  
$1 + \frac12 (D-1)(D-2)$ free fields. 

\section{Two-dimensional dilaton gravity formulation}

$D$-dimensional spacetimes with $D-2$ commuting Killing vector fields 
are described  by a line-element of the form
\be
ds^{2}=g_{\mu\nu}(x)dx^{\mu}dx^{\nu}+ g_{ab} (x) 
d\chi^a d\chi^b \quad,\label{le}
\ee
where $x^{\mu}$ are the two-dimensional coordinates and 
$\partial /\partial \chi^a$ are
the Killing vectors \cite{ver}. Let $D = n+2$, and we split the matrix 
$g_{ab}$ into
the determinant and the $SL(n,R)$ piece
\be
g_{ab}(x) = \vphi^{2/n} (x) \L_{ab} (x) 
\ee
so that $\sqrt{\det g} = \vphi (x)$ and $\det\Lambda =1$.
The corresponding Einstein equations can be derived from the following
two-dimensional action \cite{bmg}
\be
S=\int d^2x\sqrt{-g} \vphi\left[R-{1\over4}tr(\Lambda^{-1}\nabla^{\mu}\Lambda
\Lambda^{-1}\nabla_{\mu}\Lambda)\right] 
\>,\label{ai}
\ee
where $g = \det g_{\m\n}$, $R$ is a two-dimensional scalar curvature and 
$\nabla_\mu$ are covariant 
derivatives. This action describes
a two-dimensional dilaton gravity coupled to  $SL(n,R)/SO(n)$ coset
space $\sigma$-model. 

One can describe different physical solutions by appropriate
choices of the Killing vectors and the spatial topology. The 
gravitational waves correspond to the case when all Killing vectors
are spacelike and the spatial section is noncompact. When the spatial section
is compact, one has the cosmological models.  
Axisymmetric stationary solutions correspond to the case when one of the 
Killing vectors is timelike. 

The standard approach to study the dynamics of the action (\ref{ai}) is to 
fix completely the two-dimensional diffeomorphism invariance, 
so that the complete 
dynamics is contained in the Ernst equation for the matrix $\Lambda$
\be
\nabla_{\mu}(\vphi \Lambda^{-1}\nabla^{\mu}\Lambda)=0
\>.\label{ern}
\ee
For cylindrical waves,
this is done in $\vphi = r$ gauge, where $x^\mu = (t,r)$.
Note that for $n=2$, the eq. (\ref{ern}) has a duality symmetry. If 
\be
\Lambda =\frac1{\D} \left(\begin{array}{cc}h^2 + \D^2 & h\\h&1
\end{array}\right)
\> \label{p1}
\ee
is a solution then  
\be
\tilde\Lambda =\frac{\D}{r} \left(\begin{array}{cc}{\tilde h}^2 +
\frac{r^2}{\D^2} & 
\tilde h\\ \tilde h &1 \end{array}\right)
\>\label{p2}
\ee
is also a solution, provided that $\pa_\pm \tilde h =\pm \frac{r}{\D^2} 
\pa_\pm h$, where $\sqrt{2}x^\pm = t \pm r$ \cite{ernst}. This 
symmetry, which can be generalized to $n>2$ case (at least for some special 
wave polarizations, see section 3), implies that 
$d{\tilde s}\sp 2$
can have the asymptotic behavior of a flat metric in cylindrical coordinates,
i.e. when $r \to \infty$ then
\be
d{\tilde s}\sp 2 \sim -e^\g (dt\sp 2 - dr\sp 2 ) + r\sp 2 d\c_1^2 + d\c_2^2
+ d\c_3^2 + ... + d\c_n^2 \quad,\label{asy}
\ee
where $\g$ is a constant. Existence of this duality symmetry relates the
original solution to cylindrically symmetric solution, where  
coordinate $\c_1$ is the angle
of rotation around the the axis $\c_2$. We then require that the original 
solution has the asymptotics
\be
ds\sp 2 \sim -e^\g (dt\sp 2 - dr\sp 2 ) +  d\c_1^2 + d\c_2^2
+ d\c_3^2 + ... + d\c_n^2 \quad.\label{asy1}
\ee

In order to find a free-field formulation for arbitrary $n$,  we need to 
generalize the $n=2$ parametrization of the $\L$ matrix \cite{cmn}
\be
\Lambda=\left(\begin{array}{cc}e^{f}+h^2e^{-f}&e^{-f}h\\e^{-f}h&e^{-f}
\end{array}\right)
\>.
\ee
Note that for $n \ge 2$, $\L$ can be written as
\be
\Lambda = \left(\begin{array}{cc}N + h^T g h & gh\\gh& g
\end{array}\right)
\quad, \label{pn}
\ee
where $g$ is a symmetric $(n-1)\times (n-1)$ matrix, $h$ is a 
$(n-1)$-dimensional
vector (column) and $N =(\det g)^{-1}$. Given a $\L_{n-1}$, the parametrization
(\ref{pn}) implies that 
$\L_{n}$ can be written as
\be
\L_{n}=\left(\begin{array}{cc}e^{(n-1)f_{n-1}}+ e^{-f_{n-1}}h_{n-1}^T 
\L_{n-1} h_{n-1}
&e^{-f_{n-1}}\L_{n-1} h_{n-1}\\e^{-f_{n-1}}\L_{n-1} h_{n-1}&e^{-f_{n-1}}
\L_{n-1}
\end{array}\right)
\quad.\label{pn1}
\ee
The recursive relation (\ref{pn1}) gives the parametrization of $\L_n$ in
terms of $f_k$ and $h_k$ fields, where $k=1,2,...,n-1$. Note that there is
a more general parametrization given by setting $e^{f_{n-1}}= \D_{n-1}$ in 
(\ref{pn1}), where $\D_k$ can be both positive and negative (for $n=2$ this is
the parametrization (\ref{p1})). However, the asymptotics (\ref{asy1}) requires
that all $\D_k$ be positive, and hence we use the parametrization (\ref{pn1}).
 
From (\ref{pn}) it follows that
\be
\L^{-1} = \left(\begin{array}{cc}N^{-1} & -N^{-1}h\\-N^{-1}h & g^{-1}
+ N^{-1}hh^T
\end{array}\right)
\quad, \label{pn2}
\ee
so that
\be
\L^{-1}d \L = \left(\begin{array}{cc}N^{-1}(d N + h^T g d h ) & 
N^{-1}g d h\\ 
d h + (g^{-1}d g) h -N^{-1}(d N + h^T gd h)h & g^{-1}d g
- N^{-1}(gd h)h^T
\end{array}\right)
\quad. \label{pn3}
\ee
By using (\ref{pn3}) we obtain
\be
tr(\L^{-1}\pa_\mu \L \L^{-1}\pa_\nu \L ) = N^{-2} \pa_\mu N \pa_\nu N
+ 2N^{-1} \pa_\mu h^T g \pa_\nu h + tr(g^{-1}\pa_\mu g g^{-1}\pa_\nu g )
\quad.\label{rtr}
\ee
By using the recursive relation (\ref{rtr}) and the parametrization formula
(\ref{pn1}) it is easy to obtain
\be
tr(\L^{-1}\pa_\mu \L \L^{-1}\pa_\nu \L ) = \sum_{k=1}^{n-1}\left[(k^2 +k)
\pa_\mu f_k \pa_\nu f_k + 2 e^{-(k+1)f_k}(\pa_\mu h_k )^T \L_k \pa_\nu h_k 
\right]\quad.\label{ntr}
\ee

The action 
(\ref{ai}) can be now written as
\be
S =\int d^2x\sqrt{-g}\vphi\left[R-{1\over4} \sum_{k=1}^{n-1}\left[(k^2 +k)
 (\nabla f_k )^2 +2
e^{-(k+1)f_k}
(\nabla h_k )^T \L_k \nabla h_k \right] \right] \, ,\label{aiii}
\ee
which is a natural generalization of the $n=2$ action of \cite{cmn}.

The two-dimensional diffeomorphism invariance of this action implies that its
Hamiltonian form is given by
\be
S = \int dt dr \left[ \p_\r \dot\r + \p_{\vphi}\dot\vphi +
\sum_{k=1}^{n-1}\left( p_k {\dot f}_k + \p_k^T {\dot h}_k \right)
 - N_0 G_0  - N_1 G_1 \right]\,, \label{hfa}
\ee
where $\r = \log g_{11}$, and $G_0$ and $G_1$ are the constraints given by
\bea
G_0 &=&  -\p_\r \p_\vphi -\r^{\prime}\vphi^{\prime}+2\vphi^{\prime\prime} 
\nonumber\\
&+&\sum_{k=1}^{n-1} \left[{p_k^2\over (k^2 +k)\vphi}+
\frac14 (k^2 +k)\vphi(f_k^{\prime})^2 \right]\nonumber\\
&+&\sum_{k=1}^{n-1}\left[ {e^{(k+1)f_k}\over 2\phi}\p_k^T \L_k^{-1}\p_k + 
\frac12\vphi e^{-(k+1)f_k} 
(h_k^{\prime})^T \L_k(h_k^{\prime})\right]  \nonumber\\
G_1 &=& \p_{\r} {\r}^{\prime} - 2\p_\r^{\prime} + \p_\vphi \vphi^{\prime} + 
\sum_{k=1}^{n-1} \left[ p_k f_k^{\prime}+ \p_k^T h_k^{\prime}\right] \quad. 
\eea
Dots represent the $t$ derivatives and primes represent the $r$ derivatives,
and the Lagrange multipliers $N_0$ and $N_1$ are related to the components 
of the two-dimensional metric as 
\be
g_{00} = -N_0^2 + g_{11} N_1^2 \quad,\quad g_{01} = g_{11} N_1 \quad.
\ee 
The Poisson bracket algebra of the constraints is isomorphic to the 
two-dimensional diffeomorphism algebra, and hence the constraints  
generate the two-dimensional infinitesimal diffeomorphism transformations. 
The algebra of
constraints splits into a direct sum of two one-dimensional diffeomorphism 
algebras via $ C_\pm = \frac12 (G_0 \pm G_1 )$. 

\section{Equations of motion}

We will study the dynamics of the action (\ref{aiii}) in the conformal gauge
for the two-dimensional metric
\be
g_{++}= g_{--} = 0 \quad,\quad g_{+-} = - e^\r \quad,
\ee
where $\r$ is the two-dimensional conformal factor. This gauge corresponds 
to $N_0 =1$ and $N_1 = 0$ in the Hamiltonian formulation, 
and $g_{11} = e^\r$.  

Let us write the action (\ref{ai}) in
the conformal gauge
\bea
S_c &=&  \int dt dr \left[2\vphi\partial_+\partial_-\rho \right.\nonumber\\&&
\left. +
\frac12\vphi \sum_{k=1}^{n-1}\left[(k^2 +k)\pa_+ 
f_k\pa_- f_k + 2e^{-(k+1)f_k}(\pa_+ h_k )^T \L_k \pa_- h_k \right]
\right]
\,.\label{cga}
\eea
The variation of $S_c$ with respect
to $\r$ gives
\be
\partial_+\partial_-\vphi=0\>,\label{aiv}
\ee
which implies that $\vphi$ is a free field. This equation is used to completely
fix the gauge, and in our case one takes
\be
\vphi = {1\over\sqrt 2} (x^+ - x^- ) = r \quad.
\ee 
However, in order to display the free-field structure, we will take the general
solution
\be
\vphi=A_+(x^+)+A_-(x^-)
\quad,
\ee
where $A_+ (x^+ )$ and $A_- (x^- )$ are 
monotonic increasing and decreasing functions respectively,
which go as ${1\over\sqrt 2}x^+$ and $-{1\over\sqrt 2}x^- $ when 
$x^{+}\rightarrow\infty$ and $x^-\rightarrow -\infty$ respectively. 

Variation of $S_c$ with respect to $\vphi$ gives
\be
2\partial_+\partial_-\rho+ {1\over 2}\sum_{k=1}^{n-1}\left[(k^2 +k)\pa_+ 
f_k\pa_- f_k + 2e^{-(k+1)f_k}(\pa_+ h_k )^T \L_k \pa_- h_k \right] = 0\>
,\label{av}
\ee 
which can be solved as
\bea
\rho &=& a_+(x^+)+a_-(x^-) \nonumber\\
&+&{1\over4}\int_{x^+}^{\infty}dy^+ \int^{x^-}_{-\infty}dy^- 
\sum_{k=1}^{n-1}  ( k^2 +k)\pa_+ f_k\pa_- f_k \nonumber\\ 
&+&{1\over2}\int_{x^+}^{\infty}dy^+ \int^{x^-}_{-\infty}dy^- 
\sum_{k=1}^{n-1} e^{-(k+1)f_k}(\pa_+ h_k )^T \L_k \pa_- h_k 
,\label{ro}
\eea
where $a_{\pm}$ are two arbitrary chiral functions.

The non-trivial dynamics is contained in 
the Ernst equation
(\ref{ern}), which is obtained by varying the action with respect to $\L$, 
or equivalently by varying $S_c$ with respect to $f_k$ 
\bea
&&k[\partial_+(\vphi\partial_- f_k )+\partial_-(\vphi\partial_+ f_k )] +
2\vphi e^{-(k+1)f_k}(\partial_+ h_k )^T \L_k \partial_- h_k \nonumber\\
&&- {2\over k+1}\vphi\sum_{j=k+1}^{n-1}e^{-(j+1)f_j}(\partial_+ h_j )^T 
{\pa\L_j \over \pa f_k} \partial_- h_j
=0\quad,\label{avi}
\eea
and $h_k$
\bea
&&\partial_+ ( \vphi e^{-(k+1)f_k}\L_k \pa_- h_k )+\pa_- (\vphi e^{-(k+1)f_k}
\L_k\partial_+ h_k )\nonumber\\
&&-\vphi\sum_{j=k+1}^{n-1}e^{-(j+1)f_j}(\partial_+ h_j )^T 
{\pa\L_j \over \pa h_k} \partial_- h_j =0\quad.\label{avii}
\eea

In order to determine the chiral functions $a_\pm$, one needs the constraint
equations, which cannot be obtained from $S_c$, but instead one must vary
the full action with respect to $g_{++}$ and $g_{--}$ and then impose the
conformal gauge conditions. In this way one obtains
\bea
C_{\pm} &=&  \partial_{\pm}^2\vphi - \partial_{\pm}\vphi
\partial_{\pm}\rho \nonumber\\ &+& {1\over 4}\vphi
\sum_{k=1}^{n-1}\left[(k^2 +k)(\pa_\pm f_k )^2 
+ 2e^{-(k+1)f_k}(\pa_\pm h_k )^T \L_k \pa_\pm h_k \right]
=0     \quad.\label{cgc}
\eea
Imposing the constraint equations (\ref{cgc}) also requires fixing the
functions $A_\pm$, since in this way one completely fixes the 
two-dimensional 
diffeomorphism invariance. The action $S_c$ without the constraints 
(\ref{cgc}) is invariant under two-dimensional conformal transformations, 
which are two-dimensional
diffeomorphism which preserve the conformal gauge.

The $n=2$ duality symmetry can be generalized to arbitrary $n$ for special
polarizations. In the collinear polarization case, which is also
refered to as the Abelian case, we have $h_k = 0$, and a
dual solution is given by
\be
{\tilde f}_{n-1} = -f_{n-1} + {2\over n}\log r \quad,\quad 
{\tilde f}_{k} = f_{k} \quad,\quad k \le n-2 \quad.
\ee
For the case when only 
$f_{n-1}$ and $h_{n-1}$ are non-zero, a dual solution is given by
\be
{\tilde f}_{n-1} = -f_{n-1} + {2\over n}\log r \quad,\quad 
\pa_\pm {\tilde h}_{n-1} = \pm r e^{-nf_{n-1}}\pa_\pm h_{n-1} \quad.
\ee
These solutions have the asymptotics (\ref{asy}), and hence they are 
manifestly cylindrically symmetric. In general case we do not know the form 
of the dual solution with the cylindrical asymptotics, but independently of 
that the 
asymptotics (\ref{asy1}) must be satisfied, which is equivalent to requiring 
that
\be
\lim_{r\to\infty} f_k = 0 \quad,\quad \lim_{r\to\infty} h_k = 0 \quad .
\label{asy2} 
\ee
\section{Free fields}

The free-field construction of \cite{cmn} is based on the fact  
that the two-dimensional diffeomorphism
algebra of constraints admits  representations quadratic in canonical variables
\be G_0 = \frac12 \left( \eta^{ij} P_i P_j + 
\eta_{ij} Q^{i\prime } Q^{j\prime} \right) \quad,\quad 
G_1 = P_i Q^{i\prime } \quad.\label{rep}\ee
Where
$i,j=1,...,m$ and $\eta_{ij}$ is a flat Minkowskian metric.
Note that the quadratic representations of diffeomorphism constraints
in higher dimensions are not possible, since then the constraint algebra has 
structure functions which are not constants. The representation (\ref{rep})
implies that $Q^i$ are free fields and 
since (\ref{ai}) is an integrable two-dimensional theory there is a possibility
to find a canonical  transformation from the initial canonical
variables to the free-field canonical variables $(P_i,Q^i)$, 
so that the number of free fields would be $m = 1 +\frac12 n(n+1)$. 

In order to show this, we go to the conformal gauge
and insert the solutions for $\r$ and $\vphi$ into the constraints. We 
obtain 
\bea
C_{\pm}&=& \partial_{\pm}^2A_{\pm} - \partial_{\pm}A_{\pm}\partial_{\pm}
a_{\pm}\nonumber\\&&
+ \frac14 \partial_{\pm}A_{\pm}\int^{x^{\mp}}_
{\mp\infty} dy^\mp \sum_k \left[ (k^2 +k)\partial_+f_k \partial_- f_k
+2e^{-(k+1)f_k}(\partial_+ h_k)^T \L_k \partial_-h_k \right]\nonumber\\&&
+ \frac14
\vphi\sum_k \left[ (k^2 +k)\left(\partial_{\pm}f_k\right)^2+
2e^{-(k+1)f_k}(\partial_{\pm}h_k)^T \L_k \pa_\pm h_k \right]=0
\quad.\label{ci}
\eea
From $\partial_\mp C_\pm = 0$ it follows
that $\partial_{\mp}P_{\pm}=0$, where
\bea
P_{\pm}&=&
{1\over4}\partial_{\pm}A_{\pm}\int^{x^{\mp}}_{\mp\infty} dy^\mp
\sum_k \left[ (k^2 +k)\partial_+f_k \partial_-f_k
+2e^{-(k+1)f_k}(\partial_+h_k)^T \L_k \partial_-h_k \right]\nonumber\\&&
+{1\over4}
\vphi\sum_k \left[(k^2 +k)(\partial_{\pm}f_k )^2+
2e^{-(k+1)f_k}(\partial_{\pm}h)^T \L_k \pa_\pm h_k \right]
\>.\label{ep}
\eea
Since
$P_\pm$ are independent of $x^\mp$, $P_\pm$ can be evaluated by taking the 
limits $x^{\mp}\rightarrow \mp\infty$, since then the integral terms 
in (\ref{ci}) vanish. This gives
\be
P_{\pm}=\lim_{x^{\mp}\rightarrow \mp\infty} {1\over4}\vphi
\sum_k \left[ (k^2 +k)(\partial_{\pm}f_k)^2
+ 2e^{-(k+1)f_k}(\partial_{\pm}h_k)^T \L_k \pa_\pm h_k \right]\>.\label{p}
\ee

We perform a change of variables 
\be
\sqrt{{k^2 + k\over 2}} f_k\sqrt{\vphi}= \tilde F_k \quad,\quad 
h_k\sqrt{\vphi}=\tilde H_k
\ee
where $\tilde F, \tilde H$ and their derivatives are bounded in the limit
$\vphi\rightarrow\infty$. This is in agreement with
the boundary conditions (\ref{asy2}),
and the equations (\ref{avi}) and (\ref{avii}) can be written as
\be
\partial_{+}\partial_{-}\tilde F_k +O\left(1\over\sqrt{\vphi}\right)=0\>,
\ee
\be
\partial_{+}\partial_{-} \tilde H_k +O\left(1\over\sqrt{\vphi}\right)=0\>.
\ee
Therefore when $\vphi\rightarrow\infty$, one has
\be
\sqrt{{k^2 + k \over 2}}
f_k\sim {F_k\over\sqrt{\vphi}}\quad,\quad h_k\sim {H_k\over\sqrt{\vphi}}\quad,
\label{ffd}
\ee
where $F_k$ and $H_k$ are bounded free fields with bounded derivatives.

When the asymptotics (\ref{ffd}) is combined with the result (\ref{p}), we
obtain
\be
P_{\pm}={1\over2}\sum_k \left[(\partial_{\pm}F_k)^2 + 
(\partial_{\pm}H_k)^T \pa_\pm H_k \right]\quad.
\ee
If one defines
\be
X^{\pm}=A_{\pm},\qquad \Pi_{\pm}=-\partial_{\pm}a_{\pm}+{\partial_
{\pm}^2A_{\pm}\over\partial_{\pm} A_{\pm}}
\>,\label{ct1}
\ee
the constraints (\ref{cgc}) take a free-field form
\be
C_{\pm}=\Pi_{\pm}\partial_{\pm}X^{\pm}+\frac12
\sum_k \left[\left(\partial_{\pm}F_k\right)^2+(\partial_{\pm}H_k)^T 
\pa_\pm H_k\right]
\>.\label{ffc}
\ee

\section{Canonical transformation}

We have found a map from $\r,\vphi,f,h$ variables to free-field variables 
$\Pi_\pm ,X^\pm ,F_k ,H_k$
defined by (\ref{ct1}) and (\ref{ffd}), but it
is not clear whether this map represents a canonical transformation. 
Namely, although the constraints take a free-field form (\ref{ffc}), one has
to show that
\bea
&&\int_{t=const.} dr \left[ \p_\r \dot\r + \p_\vphi \dot\vphi + 
\sum_k \left(\p_{k} \dot f_k + 
(\p_{k})^T \dot h_k\right) \right]\nonumber\\
&& = \int_{t=const.}dr \left[ \Pi_+\dot X^+ +\Pi_- \dot X^- 
+\sum_k\left(P_{k}\dot F_k  +\Pi_{k} (\dot H_k)^T \right)\right] \quad,
\label{ect}
\eea
up to total time derivative. This can be shown
by examining the pre-symplectic form on the unconstrained phase space 
given by $\r,\vphi,f,h$ and their canonically conjugate momenta. This phase
space is the same as the space of solutions corresponding to the
action $S_c$ (\ref{cga}), i.e. solutions in the conformal gauge 
for which the constraint equations (\ref{cgc}) are not 
imposed, so that $A_\pm$ and $a_\pm$ are not fixed.

The unconstrained pre-symplectic one-form is defined as
\bea
\Omega &=& \int_{t=const.} dr \left[ \p_\r \d\r + \p_\vphi \d\vphi + 
\sum_k\left[\p_{k} \d f_k + (\p_{k})^T \d h_k \right]\right]\nonumber\\
&=& \int_{t=const.} dr \left[ {\pa L \over\pa \dot\r} \d\r + 
{\pa L \over \pa\dot\vphi} \d\vphi + 
\sum_k \left[{\pa L \over \pa\dot f_k} \d f_k + 
\left({\pa L \over \pa\dot h_k}\right)^T \d h_k \right] \right]\label{om}
\eea 
where $L$ is the Lagrangean density of $S_c$. This can be rewritten in the 
covariant form
\be
\Omega =
\int_{\Sigma} d\sigma_\m  {\pa L \over \pa(\pa_\m \Phi_a)} \d\Phi_a  
= \int_{\Sigma} d\sigma_\m j^\m \quad,\label{csf}
\ee
where $\delta$ stands for the exterior derivative on the space of solutions
of the equations of motion, 
$\Sigma$ is a spacelike hypersyrface and $j^\m$ is the symplectic current
one-form \cite{Witten}. Because the symplectic current is conserved 
$\pa_\m j^\m = 0$, the definition (\ref{csf})is independent of the choice of 
the hypersurface $\Sigma$.

By integrating $\pa_\m j^\m$ over the regions bounded by $t = const.$ and 
$x^\pm = \pm\infty$ hypersurfaces, one can derive  
\be
\Omega= \frac12\int_{x^-= -\infty} dx^+  j^-
+ \frac12\int_{x^+= +\infty} dx^-  j^+ 
\>.\label{sf}
\ee
The light-cone components of the one-form current
$j^{\mu}$ can be calculated from $S_c$
\be
j^+=\pa_-\vphi \delta\rho + \pa_-\r \delta\vphi - 
\frac12\vphi\sum_k \left[(k^2 +k)\partial_-f_k\delta f_k
+2 e^{-(k+1)f_k}(\partial_-h_k )^T \L_k \delta h_k \right] \>,
\ee
\be
j^-=\partial_+\vphi\delta\rho + \partial_+\r\delta\vphi
-\frac12\vphi\sum_k \left[(k^2 +k)\partial_+ f_k\delta f_k
+2 e^{-(k+1)f_k}(\partial_+ h_k )^T \L_k \delta h_k \right]\, .
\ee
By taking into account the asymptotic behavior of $\r,\vphi,f_k$ and $h_k$
for $x^\pm \to\pm\infty$ and equation (\ref{ct1}), it is easy to 
see that the symplectic two-form $\o = \d \Omega$ is given by
\bea
\omega&=&\frac12\int_{x^-=-\infty}dx^+\left[\delta X^+ \wedge\delta\Pi_+
+\sum_k\left( \delta F_{k+}\wedge
\delta \partial_+ F_{k}
+\delta H_{+k}^T\wedge\delta \partial_+ H_k \right)\right]\nonumber\\
+&\frac12&\int_{x^+=\infty}dx^-\left[\delta X^- \wedge\delta  \Pi_-
+\sum_k \left(\delta F_{k-}\wedge\delta \partial_-F_k
+\delta H_{k-}^T \wedge\delta \partial_-H_k \right)\right].
\eea
By using (\ref{sf}) we get
\be
\o =\int_{t=const.}dr \left[\delta X^+\wedge\Pi_+ + \delta X^-\wedge
\delta \Pi_- +
\sum_k\left(\delta F_k \wedge\delta \dot F_k + \delta H_k^T \wedge\delta
\dot H_k \right)\right],\label{omp}
\ee
where $F_{\pm}(x^{\pm}),H_{\pm}(x^{\pm})$ are the chiral parts of the
free fields $F$ and $H$, $(F=F_++F_-,H=H_++H_-)$. From (\ref{omp}) it follows
that
\bea
\Omega&=&\int_{t=const.} dr \left[ \p_\r \d\r + \p_\vphi \d\vphi + 
\sum_k \left(\p_{k} \d f_k + 
\p_{k}^T \d h_k\right) \right]\nonumber\\
&=& \int_{t=const.}dr \left[ \Pi_+\d X^+ +\Pi_- \d X^- 
+\sum_k\left(\dot F_{k}\d F_k  +(\dot H_{k})^T \d H_k \right)\right] \quad,
\label{oms}
\eea
which implies
\bea
&&\int_{t=const.} dr \left[ \p_\r \dot\r + \p_\vphi \dot\vphi + 
\sum_k \left(\p_{k} \dot f_k + 
\p_{k}^T \dot h_k\right) \right]\nonumber\\
&& = \int_{t=const.}dr \left[ \Pi_+\dot X^+ +\Pi_- \dot X^- 
+\sum_k\left(\dot F_{k}\dot F_k  +(\dot H_{k})^T \dot H_k \right)\right] \quad,
\label{omt}
\eea
up to a total time derivative. From (\ref{omt}) and (\ref{hfa}) it follows that
\be
S_c = \int dt dr \left[ \Pi_+\dot X^+ +\Pi_- \dot X^- 
+\sum_k\left(\dot F_{k}\dot F_k  +(\dot H_{k})^T \dot H_k \right)
- (C_+ + C_-)\right]\,, \label{sca}
\ee
where $C_\pm$ are given by (\ref{cgc}). From (\ref{sca}) it follows
that $P_k = \dot F_k$ and 
$\Pi_k = \dot H_k$, and  hence (\ref{omt}) gives (\ref{ect}). Therefore
we have a canonical transformation.

In terms of the canonical variables, (\ref{ffc}) can be written as
\be
C_{\pm}=\pm\Pi_{\pm} X^{\prime\pm}+
\frac14\sum_k\left[\left(P_{k} \pm  F_k^{\prime} \right)^2
+ \left(\Pi_{k} \pm H_k^{\prime}\right)^T 
\left(\Pi_{k} \pm H_k^{\prime}\right) \right] \>.\label{cff}
\ee
By performing a canonical transformation
\be 
2X^{\pm\prime} = \mp (\Pi_1 - \Pi_0) - X^{0\prime} - X^{1\prime} \quad,
\quad 2\Pi_{\pm} = -\Pi_0 - \Pi_1 \mp ( X^{1\prime} - X^{0\prime}) \, 
\label{ct}
\ee
the constraints take the form (\ref{rep}).

\section{Free-field expansions}

Although we have shown that a free-field formulation exists, what one really
needs are more explicit expressions for $f_k$ and $h_k$ fields in terms of 
the free fields $F_k$ and $H_k$ then the ones given by the
asymptotic relations (\ref{ffd}).

In the Abelian case the asymptotics (\ref{ffd}) is explicitly realized 
because the exact solution for the fields $f_k$ is given by 
\be
f_k =\int_{-\infty}^{\infty} d\lambda\ 
J_0\left(\lambda r \right)
\left[ A_{+k} (\lambda)e^{i\lambda t}
+A_{-k}(\lambda)
e^{-i\lambda t}\right]
\>,\label{ac}
\ee
where $J_0$ is the Bessel function and $A_{\pm k}(\lambda)$ are arbitrary 
coefficients. 
When $r\to\infty$,
this behaves as 
\bea
f_k&\sim & {1\over\sqrt{2r}}\int_{-\infty}^{\infty}d\lambda
(\pi|\lambda|)^{-{1\over2}}\left[ A_{+k}(\lambda)e^{i\lambda x^+}
e^{-i{\pi\over4}}+A_{-k}(\lambda)e^{-i\lambda x^+}e^{i{\pi\over4}}\right.
\nonumber\\ &&
\left.
+A_{+k}(\lambda)e^{i\lambda x^-}e^{-i{\pi\over4}}+A_{-k}(\lambda)
e^{-i\lambda x^-}
e^{i{\pi\over4}}\right]
\quad.
\eea
From (\ref{ffd}) it follows that
\bea
F_k &=& \int_{-\infty}^{\infty}d\lambda
(\pi|\lambda|)^{-{1\over2}}\left[ A_{+k}(\lambda)e^{i\lambda x^+}
e^{-i{\pi\over4}}+A_{-k}(\lambda)e^{-i\lambda x^+}e^{i{\pi\over4}}\right.
\nonumber\\ &&
\left.
+A_{+k}(\lambda)e^{i\lambda x^-}e^{-i{\pi\over4}}+A_{-k}(\lambda)
e^{-i\lambda x^-}
e^{i{\pi\over4}}\right] \quad,\label{aff}
\eea
and therefore $F$ is a bounded free field and $\partial_+ F,
\partial_- F,...$ are also bounded. The relations (\ref{ac}) and (\ref{aff})
give an exact relationship between $f_k$ and $F_k$, and one can obtain from 
them an exact expression for $f_k$ in terms of $F_k$.

In the non-abelian case the explicit form of the solutions is not known.
However, in $D=4$ case there is an asymptotic series expansion of $f$ and $h$ 
in terms of $F,H$ and $\vphi$ which could in principle give a non-abelian 
generalization of the expression (\ref{ac}) \cite{cmn}. The idea is to
write $f$ and $h$ as an asymptotic series expansions
\be f = {F\over \sqrt\varphi} + 
\sum_{i=1}^{\infty}{F_{(i)}\over(\sqrt\vphi)^{i+1}} 
\quad, \label{inv1}\ee
and 
\be h = {H\over \sqrt\varphi} +\sum_{i=1}^{\infty}{H_{(i)}
\over(\sqrt\vphi)^{i+1}} 
 \quad, \label{inv2}\ee
where $F_{(i)}$ and $H_{(i)}$ are functionals of $F$ and $H$. The form of these
functionals can be determined from the Ernst equations, and this can be
done explicitly because one obtains the recurrence relations
\bea
\pa_+\pa_- F_{(i)} &=& - \pa_r F_{(i-2)} - {(i-3)^2\over 16}F_{(i-4)}
+  {\cal F}_{(i-1)}
\nonumber\\
\pa_+\pa_- H_{(i)} &=&  - \pa_r H_{(i-2)} - {(i-3)^2\over 16}H_{(i-4)} + 
{\cal H}_{(i-1)}
\quad,\label{rr}
\eea 
where 
\be
{\cal F}_{(i)} = \sum_{j+\cdots +k+l+m=i}a_{lm}^{j\cdots k} F_{(j)}
\cdots F_{(k)} D_+ H_{(l)} D_- H_{(m)}  
\ee
and
\be
{\cal H}_{(i)} =\sum_{j+k=i} [D_+ F_{(j)}D_- H_{(k)} +  
D_- F_{(j)}D_+ H_{(k)}]
\quad.  
\ee
We have introduced new derivatives 
$D_\pm X_{(i)} = \pa_\pm X_{(i)} \mp {(i-1)\over 4} X_{(i-2)}$ and 
$a_{lm}^{j...k}$ are numerical coefficients, while 
$F_{(0)} = F$, $H_{(0)} =H$ and $F_{(i)} = H_{(i)} =0$ for $i < 0$. Hence all
the higher-order
$F$ and $H$ are determined from the zero-order ones.
In this way one obtains 
\be
F_{(1)} = -\frac12 H\sp 2 \quad,\quad H_{(1)} = FH \label{fdc}
\ee 
and so on. 

The same pyramid structure as (\ref{rr}) is preserved in the
$D>4$ case, and the difference is that the polynomial function 
${\cal F}_{(i)}$ depends on higher than two powers of $H$
\be
{\cal F}_{(i)} = \sum_{j+\cdots +k+l+\cdots +m+n+p =i}
a_{l\cdots mnp}^{j\cdots k} F_{(j)} \cdots F_{(k)} H_{(l)}\cdots H_{(m)} 
D_+ H_{(n)} D_- H_{(p)} \quad, \label{naf}
\ee
while ${\cal H}_{(i)}$ becomes a polynomial of order $i+2$ 
\bea
{\cal H}_{(i)} &=&\sum_{j+\cdots +k+l+\cdots +m+n+p =i}
b_{l\cdots mnp}^{j\cdots k} F_{(j)} \cdots F_{(k)} H_{(l)}\cdots H_{(l)} 
D_+ F_{(n)} D_- H_{(p)} \nonumber\\
&+&\sum_{j+\cdots +k+l+\cdots +m+n+p =i}
b_{l\cdots mnp}^{j\cdots k} F_{(j)} \cdots F_{(k)} H_{(l)}\cdots H_{(l)}
 D_- F_{(n)} D_+ H_{(p)} \nonumber\\
&+&\sum_{j+\cdots +k+l+\cdots +m+n+p =i}
c_{l\cdots mnp}^{j\cdots k} F_{(j)} \cdots F_{(k)} H_{(l)}\cdots H_{(m)} 
D_+ H_{(n)} D_- H_{(p)} \quad, \label{nah} 
\eea
where $a,b$ and $c$ are numerical coefficients. Note that
dependence on $\pa F$ and
$\pa H$ terms is always a quadratic polynomial, because the Ernst equations 
have only two derivatives. 

This structure of $\cal F$ and $\cal H$ polynomials in $D>4$ case comes from
the fact that $\L_n$ matrix is non-trivial. The matrix elements of $\L_n$ are 
polynomials in $e^{f_j}$ and $h_l$, which follows from the parametrization 
(\ref{pn1}). Furthermore, the expansions (\ref{inv1}) and (\ref{inv2}) imply
\be
\L_k = \sum_{j=0}^{\infty}\L_k^{(j)} \vphi^{-j/2} \label{la}  
\ee
where $\L_k^0$ is a constant matrix, and $\L_k^{(j)}$ for $j>0$ are polynomials
in $F_{(l)}$ and $H_{(m)}$ where $l,m \le j$. The same is valid for 
$\pa \L_k /\pa f_j$
and $\pa \L_k /\pa h_j$, so that the Ernst equations give 
(\ref{rr}) with (\ref{nah}) and (\ref{naf}). An analog of (\ref{fdc}) is
\be
F_k^{(1)} = -{1\over 2k}\left[ H_k^T H_k -{1\over k+1}\sum_{j=k+1}^{n-1}H_j^T
 H_j \right] \quad,\quad H_k^{(1)} = F_k H_k \quad.
\ee
When $H =0$, one recovers in this way the large $r$ asymptotic
expansion of the Bessel function, which is the exact solution in the Abelian
case.

\section{Conclusions}

We have shown that the $D$-dimensional cylindrically symmetric general 
relativity on spacetimes with non-compact spatial sections can be mapped 
via canonical transformation into 
a two-dimensional diffeomorphism invariant theory of $1+\frac12 (D-1)(D-2)$ 
free fields.
In the case when the spatial section is compact, which 
corresponds to cosmological models, the situation with a free-field 
realization is more complicated. In this case the simplifications 
due to $r\to\infty$ asymptotics can not be used, since $\vphi = t$ is a 
consistent gauge choice. It is clear that for $t\to\pm\infty$ one can have
asymptotic free-field solutions, but if one wants to prove
the existence of free-field variables for all times one must  
find the
exact expressions for the free fields in terms of the $D$-dimensional metric 
variables. This can be done in the $D=4$ case \cite{cmn} 
\be
(\pa_\pm F)^2 = \vphi (\pa_\pm f)^2 + \pa_\pm \vphi 
\int_{x_0^\mp}^{x^\mp}
dy^{\mp}\pa_+ f \pa_- f + 2\int_{x_0^\mp}^{x^\mp} dy^{\mp}\vphi 
e^{-2f} \pa_\pm f \pa_+ h \pa_- h ,
\label{exf}
\ee
\be
(\pa_\pm H)^2 = \vphi e^{-2f} (\pa_\pm h)^2 +  
\pa_\pm \vphi 
\int_{x_0^\mp}^{x^\mp} dy^{\mp}e^{-2f}\pa_+ h \pa_- h - 
2\int_{x_0^\mp}^{x^\mp} dy^{\mp}\vphi e^{-2f} \pa_\pm f \pa_+ h \pa_- h . 
\label{exh}
\ee
These expressions are independent of spatial topology, since they follow
from the equations of motion, which are local. In order to see whether the
transformations (\ref{exf}) and (\ref{exh}) define a canonical transformation, 
one needs to examine
the pre-symplectic form as in the section 5, but 
without using the $\vphi\to\infty$ asymptotics. Proving this and finding
a $D$-dimensional generalization of (\ref{exf}) and (\ref{exh}) is an open 
problem. Therefore, although free-fields
exist in the compact case, it is not clear whether a canonical 
transformation to free-fields exists. However, the asymptotic expansions 
(\ref{inv1}) and (\ref{inv2}) can be still used for obtaining the late-time 
solutions.  

The quantization of cylindrical gravitational waves in
terms of the free-field variables is considerably simpler than if one uses
the observables obtained from the BZM approach 
\cite{KS}. The BZM observables form  
a non-linear Yangian algebra, whose Hilbert space representations is 
difficult to find \cite{KS}. In the free-field approach the observables are 
given by the "gravitationally dressed" Fourier modes \cite{Kuchar2}, defined by
\be
F_k = {1\over2\sqrt{\pi}}\int_{-\infty}^\infty {d\l\over |\l|}
\left[ e^{i\l_\m X^\m} A_k (\l)+e^{-i\l_\m X^\m} A_k^{*} (\l)
\right]
\>,
\ee
\be
H_k = {1\over2\sqrt{\pi}}\int_{-\infty}^{\infty}
{d\l\over |\l|}\left[ e^{i\l^\m X_\m}B_k (\l)+ e^{-i\l_\m X^\m} B_k^{*} (\l)
\right]
\>,
\ee
where $\l^\m X_\m = \l_+  X^+ + \l_- X^- $ and 
$\l_\pm = \frac12 (\l \mp |\l|)$.
Upon quantization, $A(\l)$ and $B(\l)$ become creation and annihilation
operators, and the corresponding Fock space is the Hilbert space of the
quantum theory. The only subtlety in this construction is that the algebra
of the quantum constraints $C_\pm$ has an anomaly, which can be canceled
by using modified quantum constraints \cite{CJZ}
\be
{\tilde C}_{\pm}  =  C_{\pm} + 
{c\over48\pi}\left[{X^{\pm\prime\prime\prime}
\over X^{\pm\prime}}-\left({X^{\pm\prime\prime}\over
X^{\pm\prime}}\right)^2\right]
\>,
\ee
where $c$ is equal to the number of physical scalar fields, so that
$c=\frac12 (D-1)(D-2)-1$. A less straightforward task will be 
finding the
expectation values of the metric variables, since they become  complicated
functionals of the free fields. However, the expansions (\ref{inv1}) and
(\ref{inv2}) become suitable for such a task.

The Fourier modes $A (\l)$ and $B(\l)$ constitute a 
complete set of observables, and therefore the BZM observables could be in
principle expressed in terms of them. Since the BZM observables form a Yangian
$sl(n,R)$ algebra, this implies that there should be a free-field 
representation of these non-linear algebras. Also note that
for $D=4$ one can construct an affine $sl(2,R)$ algebra 
from the BZM observables, and this algebra generates
the Geroch group which is the dynamical symmetry of the theory \cite{ks2}. 
This symmetry can be easily seen in the
free-field approach, since one can construct the generators of an
affine $sl(2,R)$ algebra from $A(\l)$ and 
$B(\l)$ via the Wakimoto construction \cite{w}. Furthermore, for general
$D$ there exists a generalized Wakimoto construction
\cite{ff,bf}, which gives the generators of affine $sl(n,R)$ algebra in terms 
of $\frac12 n(n+1)-1$ creation and annihilation operators, which is exactly the
number of $A_k (\l)$ and $B_k(\l)$. This algebra will generate a 
dynamical symmetry group in the $D$-dimensional case.  

\section*{Acknowledgements} 
N.M. was partially supported by the grants PBIC/C/MAT/2150/95 and 
PRAXIS/2/2.1 /FIS/286/94 while A.M. was supported by the grant 
PRAXIS/BCC/18981/98 from the Portugese Foundation for Science and Technology.

\end{document}